\documentclass[%
onecolumn,
notitlepage,
amsmath,amssymb,
floatfix,
12pt
]{article}

\usepackage[fleqn]{amsmath}
\usepackage{amsfonts}
\usepackage{amssymb}
\usepackage{epsfig}
\usepackage{geometry}
\usepackage{ctable}
\usepackage[sort&compress,square,numbers]{natbib}

\newcommand{\cC}{{\mathcal{C}}}

\newcommand{\rV}{{\boldsymbol{r}}}

\newcommand{\Nset}{{\mathbb{N}}}
\newcommand{\Qset}{{\mathbb{Q}}}
\newcommand{\Rset}{{\mathbb{R}}}

\newcommand{\Zset}{{\mathbb{Z}}}

\def\LJ#1{$\mathrm{LJ}_{#1}$}

\begin{document}

\title{A note on the minimal pairwise distance in\\ 
       optimal Lennard-Jones $N$-body clusters$^*$\vspace{-.5truecm}}

\author{\sc{Michael K.-H. Kiessling}$^{1}$ and \sc{David J. Wales}$^2$\\
    $^1${\small{Department of Mathematics,
      Rutgers, The State University of New Jersey}}\\
     {\small{110 Frelinghuysen Rd., Piscataway, NJ 08854, USA}}\\
     {\small{email: miki@math.rutgers.edu}}\\
   $^2${\small{Yusuf Hamied Department of Chemistry, Lensfield Road, Cambridge CB2 1EW, UK}}\\
       {\small{email: dw34@cam.ac.uk}} \vspace{-.2truecm}}
% \date{Draft of August 21, 2025 \vspace{-.5truecm}}

\maketitle

\begin{abstract}
\noindent
 Good a-priori bounds on the smallest pairwise distance $r_{\mbox{\tiny{min}}}(\mbox{LJ}_N^{\rm{gmin}})$ for a three-dimensional (3D) Lennard-Jones $N$-body cluster of
 globally minimal energy can significantly reduce the computational search space in the NP-hard problem to find this configuration.
 In this contribution the virial theorem is exploited for this purpose.
 We prove that if a configuration ${\cal C}^{(N)}$ is a member of $\mbox{LJ}_N^{\rm{equ}}$ (the stationary points), then $r_{\mbox{\tiny{min}}}({\cal C}^{(N)})\leq r_{\mbox{\tiny{min}}}(\mbox{LJ}_2^{\rm{gmin}})$.
 It is also shown that if ${\cal C}^{(N)}\in$ LJ$_N^{\rm{gmin}}\subset$ LJ$_N^{\rm{equ}}$, equality holds if and only if $N\in\{2,3,4\}$.
We conjecture that $r_{\mbox{\tiny{min}}}(\mbox{LJ}_N^{\rm{gmin}}) >1$ in units for which $r_{\mbox{\tiny{min}}}(\mbox{LJ}_2^{\rm{gmin}})= 2^\frac16 \approx 1.122462048$.
 This conjectured lower bound, if correct, would improve the best lower bound currently known, $r_{\mbox{\tiny{min}}}(\mbox{LJ}_N^{\rm{gmin}})\geq 0.767764$, by about 25\%.
 In these units the smallest minimal pair distance found through numerical searches for LJ$_N^{\rm{gmin}}$ with $N\leq 1000$ % and $N\in\{309,561,923\}$ 
 is $r_{\mbox{\tiny{min}}}(\mbox{LJ}_{923}^{\rm{gmin}}) \approx  1.01361$, so the conjectured lower bound would presumably be close to optimal.
 From the virial theorem we obtain an identity for any ${\cal C}^{(N)}\in \mbox{LJ}_N^{\rm{equ}}$, which expresses $r_{\mbox{\tiny{min}}}({\cal C}^{(N)})$ in terms of the distribution of relative distances in ${\cal C}^{(N)}$.
 This result reveals interesting connections with the Erd\H{o}s distance, and related problems.
\end{abstract}

 \vfill
 \centerline{$^*$ Dedicated to Peter Schwerdtfeger on his 70th birthday.}
\bigskip
 \hrule
 \smallskip

 \copyright {\footnotesize{(2025) The authors. Reproduction, in its entirety, for non-commercial purposes is permitted.}}

% \newpage

%%%%%%%%%%%%%%%%%%%%%%%%%%
%%%%%%%%%%%%%%%%%%%%%%%%%%%%%%%%%%%%%%%%%%%%%%%%%%%%
\section{Introduction}
%%%%%%%%%%%%%%%%%%%%%%%%%%%%%%%%%%%%%%%%%%%%%%%%%%%%
%%%%%%%%%%%%%%%%%%%%%%%%%%

Since the Lennard-Jones potential was introduced, just over a century ago, it has found widespread applications in diverse fields, thanks to the realistic physical behaviour captured by a simple functional form \cite{doi:10.1021/acs.jctc.4c00135}.
New results and improvements are still being presented \cite{Schwerdtfeger2006,Smits2021},
including investigations of the lattice sums and stabilities for bulk systems 
\cite{BCPS,BurrowsPRE2021,SchwerdtfegerBurrows2022}
considered by Lennard-Jones himself.
In this article we draw attention to some intriguing geometrical aspects of $N$-body clusters with Lennard-Jones pair interactions.

Consider a configuration ${\cal C}^{(N)} := \{\rV_1, ..., \rV_N\}\subset\Rset^3$ of $N\geq 2$ distinct locations in $\Rset^3$ of classical point particles (representing atoms).
 For any pair of natural numbers $i\neq j$, each no larger than $N$, 
the Euclidean distance between the particle located at $\rV_i$ and the one at $\rV_j$ is denoted  
$|\rV_i-\rV_j|\in (0,\infty)$.

 If $r>0$ stands for the Euclidean distance of any such pair of particles under consideration, we define the Lennard-Jones interaction energy \cite{Jones1924,Jones1924a,Jones1924b,Jones1925a,JonesIngham} as
\begin{equation}\label{eq:VLJ}
V_{\mbox{\tiny{LJ}}}(r) 
:= \frac{1}{r^{12}} - \frac{1}{r^6};
\end{equation}
by a simple rescaling of distance and energy units one can recover any of the other expressions of a Lennard-Jones pair energy that are commonly used.
 The total Lennard-Jones energy for a configuration of $N\geq 2$ atoms, denoted $W({\cal C}^{(N)})$, is given by
\begin{equation}\label{W}
W({\cal C}^{(N)} ) 
: =   \sum\!\!\sum_{\hskip-.6truecm 1\leq i<j \leq N} V_{\mbox{\tiny{LJ}}}(|\rV_i-\rV_j|) .
\end{equation}

 A configuration ${\cal C}^{(N)}$ may or may not evolve in time under Newtonian dynamics, but as long as the $N$ atoms remain in a bounded region for all time (after at most transforming to a Galilei frame in which the total momentum of the configuration vanishes) one speaks of a bound $N$-body cluster. 
 We denote the family of all $N$-body clusters bound by Lennard-Jones interactions as LJ$_N$.

 Here we are interested in the subsets LJ$_N^{\rm{equ}}$ and LJ$_N^{\rm{gmin}}$ of LJ$_N$, where LJ$_N^{\rm{gmin}}$ denotes the global minimizers of $W({\cal C}^{(N)})$, while LJ$_N^{\rm{equ}}$ denotes all stationary points of $W({\cal C}^{(N)})$; the latter are force equilibria. 
 It is easy to show that for each $N\geq 2$ there exists a configuration 
${\cal C}^{(N)}_{\rm gmin}$, the {\it optimal $N$-particle Lennard-Jones cluster}, which 
minimizes the total Lennard-Jones energy $W({\cal C}^{(N)})$ globally over the set of all $N$-point configurations.
Thus, LJ$_N^{\rm{gmin}}$ is not empty when $N\geq 2$, and therefore 
LJ$_N^{\rm{equ}}$ is not empty a forteriori.

The energy of any configuration is invariant to all permutations of identical particles, inversion through the origin, and overall translation and rotation.
 Structures that differ only by an overall translation or rotation are easily identified; 
 rotational alignment can be achieved using Lagrange multipliers \cite{Kabsch78} or quaternions \cite{Kearsley89,CoutsiasSD04}.
Usually also the permutation-inversion isomers of each stationary point are lumped together for sampling purposes, since the number of distinct structures is known analytically from the point group \cite{FrenkelW11,C5SM01014D,Wales03}.
The GMIN, OPTIM, and PATHSAMPLE programs for exploring energy landscapes employ an iterative scheme to perform optimal alignment between different local minima \cite{WalesC12}, based on the shortest augmenting path algorithm \cite{jonkerv87} for the permutational part.
This analysis provides the shortest Euclidean distance between permutational isomers of different structures, while a standard orientation scheme identifies permutation-inversion isomers in an initial filter \cite{WalesC12,GriffithsNW17}.
For the present application, alignment is not an issue, and we can use any permutation-inversion isomer of the global minimum.

 For small enough $N$ the set LJ$_N^{\rm{gmin}}$ has only one element (aside from permutation-inversion isomers with the same energy), but we are not aware of a proof of uniqueness for global minimum Lennard-Jones clusters in 3D with general $N$.
 In contrast, LJ$_N^{\rm{equ}}$, the set of all stationary points, has almost always more than one element, and is expected to increase exponentially in size with $N$ \cite{StillingerW82,StillingerW84,WalesD03}.
 The only exception, when $D=3$, is when $N=2$; it is readily seen that LJ$_2^{\rm{equ}}=$ LJ$_2^{\rm{gmin}}$. 
 Only when $D=1$ is LJ$_N^{\rm{equ}}=$ LJ$_N^{\rm{gmin}}$ for all $N\geq 2$.

 Focusing on the 3D problem, we recall that for $N\in\{2,3,4,5\}$ the optimal structure LJ$_N^{\rm{gmin}}$ can be established rigorously, but for larger $N$ the problem requires numerical searches on a computer, and even with the fastest available algorithms the problem becomes increasingly difficult as $N$ increases because LJ$_N^{\rm{min}}$, the set of all local energy minimizing Lennard-Jones clusters, appears to grow exponentially fast with $N$ \cite{StillingerW82,StillingerW84,WalesD03}, as mentioned above.
 In fact, the problem of finding  LJ$_N^{\rm{gmin}}$ has been proved to be NP-hard \cite{WilleVennik}.

 In the absence of a sure-fire algorithm that locates LJ$_N^{\rm{gmin}}$ in polynomial time, probabilistic searches are used. 
 Good a-priori bounds on the search space are desirable to eliminate as many irrelevant configurations as possible.
 One important quantity that bounds the size of the search space is the minimal pairwise distance $r_{\mbox{\tiny{min}}}(\mbox{LJ}_N^{\rm{gmin}})$ that can occur in any $N$-body cluster global minimum.
  It has been shown first in \cite{Xue} that there is an $N$-independent nontrivial lower bound to $r_{\mbox{\tiny{min}}}(\mbox{LJ}_N^{\rm{gmin}})$, namely (rescaled into our units)  $r_{\mbox{\tiny{min}}}(\mbox{LJ}_N^{\rm{gmin}})\geq 1/2^\frac{5}{6}\approx 0.561231$.
 In subsequent papers, first \cite{Vinko}, % $r_{\mbox{\tiny{min}}} \geq 0.6945$ 
then \cite{SchachingerETal}, %  $r_{\mbox{\tiny{min}}} \geq 0.7631$
and more recently \cite{Yuhjtman}, the theoretical lower bound has been improved, with the currently best estimate $r_{\mbox{\tiny{min}}}(\mbox{LJ}_N^{\rm{gmin}})\geq 0.767764$ \cite{Yuhjtman}.
 The important question is: What is the optimal lower bound?

%%%%%%%%%%%%%%%%%%%%%%%%%%
%%%%%%%%%%%%%%%%%%%%%%%%%%%%%%%%%%%%%%%%%%%%%%%%%%%%
\section{The minimal pair distance in LJ$_N^{\mbox{\small{equ}}}$ and in LJ$_N^{\mbox{\small{gmin}}}$}
%%%%%%%%%%%%%%%%%%%%%%%%%%%%%%%%%%%%%%%%%%%%%%%%%%%%
%%%%%%%%%%%%%%%%%%%%%%%%%%
 
 For $N\geq 2$ the minimal distance among all pairs of particles in an 
equilibrium configuration ${\cal C}^{(N)}\in$ LJ$_N^{\rm{equ}}$ is defined as
\begin{equation}
r_{\mbox{\tiny{min}}}({\cal C}^{(N)}) := \min_{\{\rV_i\neq \rV_j\} \subset {\cal C}^{(N)}} |\rV_i-\rV_j|.
\end{equation}
 In particular, when $N=2$ the equilibrium problem is trivial, as manifestly there is only a single equilibrium configuration, which therefore is the global energy minimizer; i.e., $\{{\cal C}^{(2)}_{\rm gmin}\}=$ LJ$_2^{\rm equ}$, and
the particles in ${\cal C}^{(2)}_{\rm gmin}$ are separated by the distance
\begin{equation}
r_{\mbox{\tiny{min}}}^{}({\cal C}^{(2)}_{\rm gmin}) 
= 2^\frac16 \approx 1.122462048.
\end{equation}
 This dimer equilibrium configuration can be used as a building block to construct the global minima for $N\in\{3,4\}$, namely the equilateral triangle and the regular tetrahedral configurations, in which all pairwise distances are the same, and equal to $r_{\mbox{\tiny{min}}}^{}({\cal C}^{(2)}_{\rm gmin})$.
Thus, if ${\cal C}^{(N)}\in$ LJ$_N^{\rm gmin}$, then
\begin{equation}
\qquad
 r_{\mbox{\tiny{min}}}^{}({\cal C}^{(N)})
 = 2^\frac16 \quad\mbox{for} \quad N\in\{2,3,4\}.
\end{equation}

 It is an elementary geometrical fact that these global LJ energy minimizers for $N\in\{2,3,4\}$ are the only LJ equilibrium configurations for which all pairs of particles in the configuration exhibit the same distance, in fact the same distance as in the LJ$_2^{\rm gmin}$.
 We will use this fact to show that these are the only three LJ equilibrium configurations in $\Rset^3$ for which the $N=2$ equilibrium distance is the minimal pairwise distance.

A minimal interatomic distance below the pair equilibrium value of $2^{1/6}$ is expected for structures based on icosahedra, since the radial distance for a regular icosahedron is $\sqrt{1+\phi^2}/2\approx0.951$ times the edge length, where $\phi=(\sqrt{5}+1)/2$, and $2^{1/6}\sqrt{1+\phi^2}/2\approx1.0675$.
We will prove that the dimer equilibrium value of $2^{1/6}$ is in fact a strict upper bound on $r_{\mbox{\tiny{min}}}^{}({\cal C}^{(N)})$ for all ${\cal C}^{(N)}\in$ LJ$_N^{\rm equ}$ with $N>4$.

For our proofs we will invoke the virial theorem.
\medskip

\subsection{The virial identity}

For ${\cal C}^{(N)}\in$ LJ$_N^{\rm equ}$ we write ${\cal C}^{(N)}
=: \{\rV_1^{\mbox{\tiny{equ}}},...,\rV_N^{\mbox{\tiny{equ}}}\}$.
 For such Lennard-Jones equilibrium configurations (stationary points) the virial theorem yields 
\begin{equation}
2\, \sum\!\!\sum_{\hskip-.6truecm 1\leq i<j \leq N} 
\frac{1}{|\rV_i^{\mbox{\tiny{equ}}}-\rV_j^{\mbox{\tiny{equ}}}|^{12}}
=
 \sum\!\!\sum_{\hskip-.6truecm 1\leq i<j \leq N} \frac{1}{|\rV_i^{\mbox{\tiny{equ}}}-\rV_j^{\mbox{\tiny{equ}}}|^{6}}.
\end{equation}
 Since  $2 = r_{\mbox{\tiny{min}}}({\cal C}^{(2)}_{\rm gmin})^6$, by homogeneity 
the virial theorem yields for all $N$-particle Lennard-Jones equilibrium configurations the identity
\begin{equation}\label{Virial}
r_{\mbox{\tiny{min}}}({\cal C}^{(N)}) 
= r_{\mbox{\tiny{min}}}({\cal C}^{(2)})
 \left(\frac{R({\cal C}^{(N)} ) }{A({\cal C}^{(N)}) } \right)^{\frac16},
\end{equation}
where ${\cal C}^{(2)} = {\cal C}^{(2)}_{\rm gmin}$, 
\begin{equation}\label{R}
R({\cal C}^{(N)}) 
: =   \sum\!\!\sum_{\hskip-.6truecm 1\leq i<j \leq N} 
\left(\frac{r_{\mbox{\tiny{min}}}({\cal C}^{(N)})}
           {|\rV_i^{\mbox{\tiny{equ}}}-\rV_j^{\mbox{\tiny{equ}}}|}\right)^{12}
\end{equation}
and
\begin{equation}\label{A}
A({\cal C}^{(N)} ) 
: =   \sum\!\!\sum_{\hskip-.6truecm 1\leq i<j \leq N} 
\left(\frac{r_{\mbox{\tiny{min}}}({\cal C}^{(N)})}
           {|\rV_i^{\mbox{\tiny{equ}}}-\rV_j^{\mbox{\tiny{equ}}}|}\right)^{6}.
\end{equation}
 Note that (\ref{Virial}) is simply a rewriting of the virial identity.
 Indeed, since
$r_{\mbox{\tiny{min}}}({\cal C}^{(N)})$ can be cancelled from left and right side in (\ref{Virial}),
one may wonder how this result provides insight into $r_{\mbox{\tiny{min}}}({\cal C}^{(N)})$?
We now show how it is useful.

\subsection{Upper bound on $r_{\mbox{\tiny{min}}}({\cal C}^{(N)})$ when ${\cal C}^{(N)}\in$ LJ$_N^{\rm{equ}}$}

 Since $r_{\mbox{\tiny{min}}}({\cal C}^{(N)})\leq |\rV_i-\rV_j|$  for 
$\{\rV_i\neq \rV_j\}\subset {\cal C}^{(N)}\in$ LJ$_N^{\rm{equ}}$,
and since $x^{12} \leq x^6$ for $0<x\leq 1$, it follows that $R({\cal C}^{(N)}) \leq A({\cal C}^{(N)})$, 
and so, for all $N\geq 2$, and any equilibrium configuration ${\cal C}^{(N)}$, we have
\begin{equation}
r_{\mbox{\tiny{min}}}^{}({\cal C}^{(N)}) \leq r_{\mbox{\tiny{min}}}^{}({\cal C}^{(2)}_{\rm gmin}).
\end{equation}
 Furthermore, this $N$-independent upper bound can be attained if and only if the distances between the particles in all pairs
$\{\rV_i\neq \rV_j\}\subset {\cal C}^{(N)}$  are identical, and therefore identical to $r_{\mbox{\tiny{min}}}({\cal C}^{(N)})$,
in which case $A({\cal C}^{(N)}) = R({\cal C}^{(N)})$.
 We already know that this is the case for the global minima LJ$_N^{\rm{gmin}}$ when $N\in\{2,3,4\}$.
 Moreover, we already noted the elementary geometrical fact that when $N>4$, then 
there are no $N$ point configurations in $\Rset^3$ in which all pairwise distances are the same.
 However, when a distance $|\rV_i-\rV_j| \neq r_{\mbox{\tiny{min}}}^{}({\cal C}^{(N)})$ for at least one pair $(i\neq j)$, 
then  ${r_{\mbox{\tiny{min}}}({\cal C}^{(N)})}/ {|\rV_i^{\mbox{\tiny{equ}}}-\rV_j^{\mbox{\tiny{equ}}}|}< 1$ for that
pair, and since $x^{12} < x^6$ when $x\in(0,1)$, we then have strict inequality 
$R({\cal C}^{(N)}) < A({\cal C}^{(N)})$, which holds for any equilibrium configuration with $N>4$.

 Thus we have proved the following theorem.
\bigskip

\noindent
{\bf Theorem}: \textit{Consider the family $\left\{{\cal C}^{(N)}\right\}^{}_{N\geq 2}$ of stationary points
of the Lennard-Jones energy given in equation (\ref{W}), i.e.~consider all $N$-particle LJ equilibrium configurations.
 Then
\begin{equation}
 \max_{N\geq 2}  \min_{\{\rV_i\neq \rV_j\} \subset {\cal C}^{(N)}} |\rV_i-\rV_j| = 2^\frac16,
\end{equation}
and this maximum is attained if and only if} ${\cal C}^{(N)}\in$ LJ$_N^{\rm{gmin}}$ \textit{with $N\in\{2,3,4\}$.}
\bigskip

\subsection{\hspace{-2pt}Conjectured lower bound for $r_{\mbox{\tiny{min}}}({\cal C}^{(N)})$ when ${\cal C}^{(N)}\in$ LJ$_N^{\rm{{gmin}}}$}

There is empirical evidence and theoretical plausibility for the
\medskip

\noindent
{\bf Conjecture}: \textit{For all} $N\geq 2$, \textit{if} ${\cal C}^{(N)}\in$ LJ$_N^{\rm{{gmin}}}$ \textit{in $\Rset^3$, then}
\begin{equation}
r_{\mbox{\tiny{min}}}({\cal C}^{(N)}) \geq 1. 
\end{equation} 

\begin{figure}[p]
\centering
\includegraphics[width=1.0\textwidth]{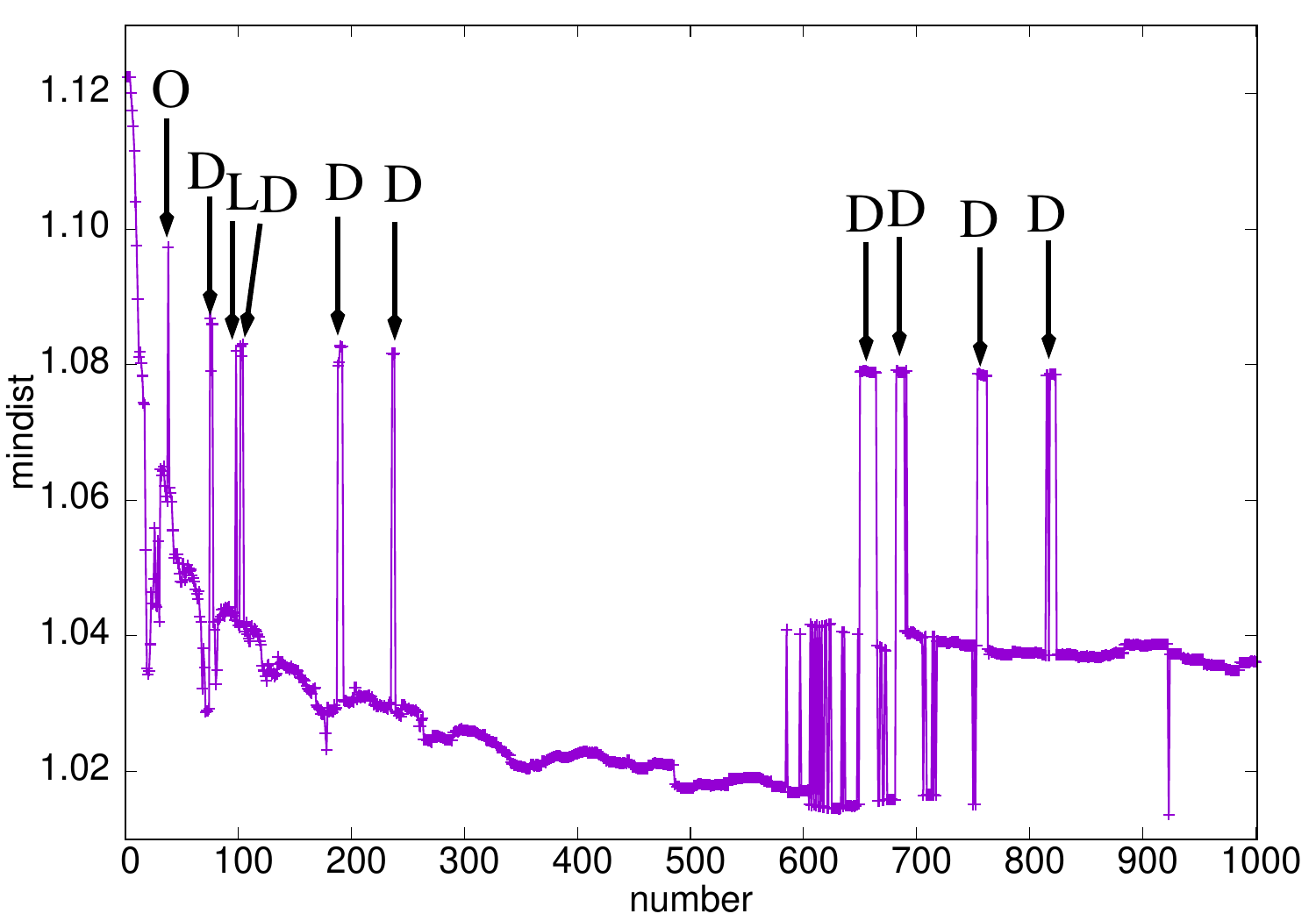}
 \caption{Minimum distance in Lennard-Jones clusters with globally minimal energy in the size range from 2 to 1000 atoms.
The points are joined to guide the eye.
Structures based on the truncated cuboctahedron (O), the Leary tetrahedron (L) \cite{Leary}, and Marks decahedra (D) \cite{marks84} are highlighted. In each case the minimum distance is significantly larger than for the neighbouring clusters based on Mackay icosahedra \cite{mackay62}.
 \label{fig:mindist}}\vspace{-.4truecm}
\end{figure}

 As to the empirical evidence for this Conjecture:
 Inspecting first the list of putative LJ$_N$ global energy minimizers at \cite{CCD}, which covers all $N\leq 150$, and then enlarging this list to all $N\leq 1000$, we find that the smallest minimal distance $r_{\mbox{\tiny{min}}}^{}({\cal C}^{(N)}_{\rm gmin})$ occurs for $N=923$, with $r_{\mbox{\tiny{min}}}(\mbox{LJ}_{923}^{\rm{gmin}}) \approx  1.01361$; see Figure \ref{fig:mindist}.
 % $r_{\mbox{\tiny{min}}}^{}({\cal C}^{(71)}_{\rm gmin}) 
 % \approx 1.028758632$

Figure \ref{fig:mindist} shows the minimum distance in the \LJ{N} clusters with lowest known energy as a function of size up to 1000 atoms.
 The largest minimal pair distance occurs for $N\in\{2,3,4\}$, as proved earlier, and the smallest minimal distance observed in this range of $N$ values occurs at $N=923$, a centered Mackay icosahedron \cite{mackay62}.
 All the distances in this figure are larger than 1.01, supporting our Conjecture.
 More important than the fact that $r_{\mbox{\tiny{min}}}({\cal C}^{(N)}_{\rm gmin})>1$ for all $2\leq N\leq 1000$ is the following observation:
 Although the range of $N$ values is relatively small in this plot, Figure \ref{fig:mindist} visibly suggests that the monotonic decreasing (though not strictly decreasing) function $N\mapsto \min_{n\leq N}r_{\mbox{\tiny{min}}}({\cal C}^{(n)}_{\rm gmin})$, which converges to the smallest possible minimal distance of any globally energy-minimizing LJ$_N$ cluster, may well converge to a value larger than 1.
Of course this is only suggestive and does not prove that there is no value of $ r_{\mbox{\tiny{min}}}^{}({\cal C}^{(N)}_{\rm gmin})$ smaller than 1 for some $N$ other than those inspected.

{The plot also reveals some interesting, highly nonmonotonic, structure, which is related to the underlying cluster morphology. 
For larger $N$ we see minimum distances around 1.02, 1.04, and 1.08. 
The smallest value corresponds to a distance involving the centre atom in structures based on a Mackay icosahedron \cite{mackay62}, where there is a central 13-atom icosahedral motif.
Inspection of selected structures reveals that minimum distances around 1.04 come from icosahedral-based geometries where the centre atom is missing.
The minimum distance occurs for two atoms in the hollow 12-atom (approximately) icosahedral core.
Values around 1.08 come from the distance between the centre atom and a first shell neighbour in decahedral structures \cite{Ino69,marks84}. 
Here, the central motif is a centred 13-atom Ino decahedron \cite{Ino69}.
These structural effects can be related to strain energy \cite{doyew95a,doyew96d}, which makes icosahedral packing unfavourable for larger clusters \cite{XiangCCS04}.
In fact, the compression around the central atom is reflected in the generally decreasing distance with $N$ for the centred icosahedral structures, until we enter a size regime where alternative packings become more favourable \cite{DoyeC02}.
Crystal lattice studies suggest that a a minimal distance value of $\approx 1.09$ is expected for very large $N$ values.
However, before this limit is reached we expect to see structural competition punctuated by nonmonotonic excursions at particular sizes.
Entropic effects will also play a role at finite temperature, and have been investigated in previous work \cite{DoyeC02,NoyaD06}.
The vast majority of global minima up to $N=1000$ can be described in terms of icosahedral packing schemes with Mackay or anti-Mackay overlayers \cite{doyew97a,XJCSinJCP2004,XiangCCS04}. 
The known exceptions are the truncated octahedron ($N=38$), the Leary tetrahedron \cite{Leary} ($N=98$) and structures based on Marks decahedra \cite{marks84,ROMERO199987,XiangCCS04} ($N=75-77,\ 102-104,\ 187-192,\ 236-238,\ 650-664,\ 682-691,\ 754-762,\ 815-823$). 
These discontinuities in packing all produce features at larger values of the minimum distance in Figure \ref{fig:mindist}.}

 Back to our Conjecture, in \cite{KW} we noted that there is also theoretical
evidence from studies of crystal structures in the thermodynamic limit
$N\to\infty$, which we repeat here:
 In \cite{JonesIngham} the minimal pair distance was computed for spatially unbounded simple cubic, bcc and fcc Lennard-Jones crystals, by minimizing their energy per monomer.
 They found that the fcc crystal has the lowest energy per particle among these three regular lattices, and for a long time it was thought that fcc is  the optimal \textcolor{black}{crystal structure} in the limit $N\to\infty$.
 Now we know that fcc is the optimal \emph{standard lattice} structure, with a minimal pair distance that can then be computed from the results in \cite{JonesIngham} as 
\begin{equation}
 r_{\mbox{\tiny{min}}}^{\mbox{\tiny{fcc}}} \approx 1.090172.
\end{equation}
 However, recently it was found empirically  \cite{ZschornakETal}, and then proved rigorously \cite{BST}, that the hcp structure is the true crystalline ground state configuration in the thermodynamic limit.
 Note that the hcp crystal is not a regular lattice in the same sense as fcc, bcc, and simple cubic crystals, because a two atom basis is required for the repeated unit, with two atomic environments.
 Using the results of \cite{BCPS}, the minimal pair distance in the optimal crystal structure can be computed to be
\begin{equation}
 r_{\mbox{\tiny{min}}}^{\mbox{\tiny{hcp}}}
 \approx 1.090167,
\end{equation}
which is just a tiny bit smaller than $ r_{\mbox{\tiny{min}}}^{\mbox{\tiny{fcc}}}$.
 It is clear that $ r_{\mbox{\tiny{min}}}^{\mbox{\tiny{hcp}}} >  r_{\mbox{\tiny{min}}}^{}({\cal C}^{(923)}_{\rm gmin})$.

Numerical simulations suggest that with growing $N$ the LJ$_N$ global minimum configurations do converge (on nested compact subsets of $\Rset^3$) to an hcp lattice structure (after for a long while fcc seemed to be the attractor). 
 If proven true, this then means that $r_{\mbox{\tiny{min}}}^{}({\cal C}^{(N)}_{\rm gmin})$ has to converge to $\approx 1.09$ (see above). 
 This in turn would prove that eventually, for large enough $N$, no minimal pairwise distance will be smaller than the conjectured smallest possible pairwise distance. 
 A counterexample would therefore have to be found below some ``large enough $N$.''
 
 We also emphasize that we are considering the global minimum energy configurations, and smaller distances could occur in local minima, or in a saddle point configuration.
 These possibilities are interesting, but beyond the scope of the present note.
However, we have checked the minimum distances for some local minima in the size range up to \LJ{1000}
and confirmed that some slightly smaller pair distances do occur.

{We note that for the sequence of centered Mackay icosahedra the minimal pairwise distance $r_{\rm min}(N)$ decreases monotonically when $N\in\{13, 55, 147, 309, 561, 923\}$.
  Since the strain of icosahedral packing increases with size we would expect the minimal distance to decrease further, so long as Mackay icosahedra are still local minima.
  This is a purely geometric effect, independent of whether the Mackay icosahedron is the global minimum.
 In fact, numerical experiments \cite{DoyeC02} show that the strain energy for isosahedral packing will probably make fcc packing more favourable around $N\approx 2\times 10^5$. The global minima based on icosahedral packing lack the centre atom for the larger clusters in Figure \ref{fig:mindist}. Hence, centred icosahedra can only be local minima at best in this size range, and their minimal distance is then not relevant for our Conjecture.}

\medskip

 In the following we present a plausible, though not conclusive, strategy that could prove our Conjecture.
 Consider the Baxter sticky hard sphere model with spheres of radius $1/2$. 
 Then any minimal energy configuration is a configuration with the maximal number of contact points (which may not be unique), and thus the minimal pairwise distance in such an $N$-body cluster made of sticky hard spheres equals 1, independent of $N$.
 Now pick any $N\geq 2$ and choose a minimal energy sticky hard sphere configuration as initial configuration for a Lennard-Jones energy minimizing gradient flow dynamics (or gradient-based minimisation algorithm), viz.
\begin{equation}\label{LJgradFLOW}
\forall i\in \{1,...,N\}:\ \dot\rV_i(t) = - \nabla_i W\big({\cal C}^{(N)}\big)\Big|^{}_{{\cal C}^{(N)}(t)}.
\end{equation}
 It is generally expected, empirically tested, but not rigorously proved,
% (to my knowledge --- MK), 
that at least one of the optimal sticky hard sphere configurations will be evolved by the gradient flow toward a global minimum ${\cal C}^{(N)}_{\rm gmin}\in$ LJ$_N^{\rm gmin}$ of the Lennard-Jones potential. 
 Assuming that this scenario holds, it is easy to see that any two initially touching spheres will experience a repulsive pairwise LJ force initially, which will act to increase their pairwise distance. 
 Of course, there will generally be additional forces exerted on these two particles by the others, and when located favorably they may initially partly or completely cancel the repulsive force between the given pair, or worse, overpower it so that the two particles move even closer to each other (temporarily at least).

 Here is an example that this worst case scenario can happen in 1D: Consider a constrained minimal energy configuration of four sticky hard spheres of radius 1/2 each, constrained to a line.
 The configuration is a chain with centers at $\{\frac{-3}{2},\frac{-1}{2},\frac12,\frac32\}$ (up to overall translation or rotation).
 With this initial configuration the LJ flow will initially move the two outer particles further out, and the two inner ones closer together; this is easily proved rigorously.
 Hence the separation of the two inner particles becomes $<1$ in the earliest phase of the evolution.
 Running the LJ flow reveals that eventually the separation of the two inner particles increases to a distance $>1$, yet smaller than the outer nearest neighbor distance.

 Now, such a linear chain arrangement of the four spheres is not an energy-minimizing sticky-hard-sphere configuration in $\Rset^3$ (a regular tetrahedron is), and it is easy to see by symmetry that the regular tetrahedral initial configuration with pairwise distances all equal to 1 will uniformly and monotonically expand under the LJ flow that preserves the regular tetrahedral shape of the configuration. 

 Inspecting the $N=5$ LJ flow starting from the energy-minimizing sticky-hard-sphere configuration in $\Rset^3$ also shows an expansion of all distances {\it initially}, but it is not uniform (due to lower symmetry than for $N=4$), yet after a while one of the three different distances involved begins to shrink while the two others continue to expand. 
 This reversal from expansion to shrinking means that one cannot hope to prove a monotonic expansion for all times of all distances under an LJ flow that starts from an energy-minimizing sticky-hard-sphere configuration.
 However, in the evolving $N=5$ configurations the smallest distance at a finite time remains the smallest distance for all time, and it increases monotonically to a value $>1$. 
 If this observation is generic, then our Conjecture is true, and would be provable by demonstrating the monotonic increase of the smallest pairwise distance in an LJ flow that starts with a sticky hard sphere configuration of global minimal energy in $\Rset^3$.

 This ends our theoretical plausibility reasoning in support of our Conjecture.

  Our Conjecture, if true, may be difficult to prove.
  In the meantime, it is certainly desirable to improve the best rigorous lower bound on  $r_{\mbox{\tiny{min}}}^{}({\cal C}^{(N)}_{\rm gmin})$ known so far, which differs by more than 25$\%$ from the suspected optimal lower bound.
 This observation suggests that there is ample room for improvement.

\medskip

\subsection{\hspace{-7pt}Connection of $r_{\mbox{\tiny{min}}}({\cal C}^{(N)}_{\rm equ})$ with the distance spectrum of ${\cal C}^{(N)}_{\rm equ}$}

 For any $N$-particle cluster ${\cal C}^{(N)}$ there are $\genfrac{(}{)}{0pt}{1}{N}{2}$ different ways of choosing a pair of two distinct particles in the cluster.
 {It is easy to see that there can be as many distinct pairwise distances in an $N$-point configuration; for instance, placing $N$ point particles on the first $N$ locations of the sequence $\{0,1,3,7,12,...\}\subset\Zset$ (which is sequence A025582 at OEIS \cite{OEIS}) accomplishes this}.
 Yet in any particular Lennard-Jones equilibrium cluster ${\cal C}^{(N)}\in$ LJ$_N^{\rm equ}$ there typically are fewer distinct distances, because LJ stationary points often possess symmetry.
 The obvious exception occurs when $N=2$, since $\genfrac{(}{)}{0pt}{1}{2}{2}$ is equal to $1$, trivially.

 Let ${\cal C}^{(N)}$ be an LJ$_N$ cluster stationary point, and let $\nu({\cal C}^{(N)})$ denote the number of different distances in this cluster, which we may abbreviate as $\nu(N)$ when the context is clear.
 We note that for the global energy-minimizing $N$-particle Lennard-Jones clusters LJ$_N^{\rm gmin}$ one has $\nu(N) = 1$ if $N\in\{2,3,4\}$, yet $\nu(N) \in\{2,..., \genfrac{(}{)}{0pt}{1}{N}{2}\}$ if $N>4$, while with general LJ equilibria $\nu(N)>1$ is feasible already when $N>2$. 

 If ${\cal C}^{(N)}\in$ LJ$_N^{\rm equ}$, we now abbreviate the minimal distance by $r_{\mbox{\tiny{min}}}^{}({\cal C}^{(N)}) =: r_1(N)$, and the next smallest pairwise distance in ${\cal C}^{(N)}$ by $r_2(N)$, and so on. 
 We introduce the ratios 
\begin{equation}
\frac{r_1(N)}{r_k(N)} := \rho_k(N) \in (0,1], \quad k = 1,...,\nu(N);
\end{equation}
note that $\rho_1(N)=1$ for all $N\geq 2$, and that $\rho_k(N)<1$ if $\nu(N)>1$ and $k>1$. 
 We also introduce 
\begin{equation}
C_k(N) := \Big| \{ \mbox{distinct\ pairs\ with\ ratio}\ \rho_k(N)\}\Big|\, \in \Nset;
\end{equation}
the allowed values of $C_k(N)$ (orbit sizes) are determined by subgroups of the point group for configuration ${\cal C}^{(N)}$, and  
\begin{equation}
\sum_{k=1}^{\nu(N)} C_k(N) =  \genfrac{(}{)}{0pt}{1}{N}{2}.
\end{equation}
 Then for any LJ equilibrium cluster ${\cal C}^{(N)}$ we can rewrite the ratio ${R({\cal C}^{(N)})}/{A({\cal C}^{(N)})}$ as follows,
\begin{equation}\label{RoverAa}
\frac{R({\cal C}^{(N)})}{A({\cal C}^{(N)})} = 
\frac{\sum\limits_{k=1}^{\nu(N)} C_k(N) \rho_k(N)^{12}}{\sum\limits_{k=1}^{\nu(N)} C_k(N) \rho_k(N)^{6}}.
\end{equation}
 A final reduction is achieved by using the fact that $C_1\in\Nset$ and $\rho_1 = 1$, so we can rewrite (\ref{RoverAa}) as
\begin{equation}\label{RoverAb}
\frac{R({\cal C}^{(N)})}{A({\cal C}^{(N)})} = 
\frac{1+\sum\limits_{k=2}^{\nu(N)} c_k(N)\rho_k(N)^{12}}{1+\sum\limits_{k=2}^{\nu(N)}c_k(N)\rho_k(N)^{6}},
\end{equation}
where $c_k := C_k/C_1$.
 Thus, to compute the ratio ${R({\cal C}^{(N)})}/{A({\cal C}^{(N)})}$ only the number count ratios $C_k/C_1\in \Qset$ are needed, not the actual number counts $C_k$, in addition to the distance ratios $\rho_k<1$, for $k\in\{2,...,\nu(N)\}$.
\medskip

\subsection{The Conjecture revisited}

 Recalling the virial identity (\ref{Virial}), our rewriting of ${R({\cal C}^{(N)})}/{A({\cal C}^{(N)})}$ as (\ref{RoverAb}) for any ${\cal C}^{(N)}\in$ LJ$_N^{\rm equ}$ has an interesting application for the problem of estimating the minimal pairwise distance in any global minimum Lennard-Jones $N$-body cluster.
 Namely, our conjectured lower bound on $r_{\mbox{\tiny{min}}}^{}({\cal C}^{(N)})$ for when ${\cal C}^{(N)}\in$ LJ$_N^{\rm gmin}$, which reads $r_{\mbox{\tiny{min}}}^{}({\cal C}^{(N)})\geq 1$ in units where $r_{\mbox{\tiny{min}}}^{}({\cal C}^{(2)}) =2^\frac16$, would be true if one could show that ${R({\cal C}^{(N)})}/{A({\cal C}^{(N)})} \geq 1/2$ for all ${\cal C}^{(N)}\in$ LJ$_N^{\rm gmin}$.
 Equivalently, the estimate ${R({\cal C}^{(N)})}/{A({\cal C}^{(N)})} \geq 1/2$ for all ${\cal C}^{(N)}\in$ LJ$_N^{\rm gmin}$ would imply that ${r_{\mbox{\tiny{min}}}^{}({\cal C}^{(N)})}/{r_{\mbox{\tiny{min}}}^{}({\cal C}^{(2)})}\geq {1}/{2^\frac16}$ 
 in any units of energy and distance for these Lennard-Jones global minima.

 To obtain good upper and lower estimates on the $\rho_k(N)$ and the $c_k(N)$, and on $\nu(N)$, may not be any easier than obtaining good lower estimates on $r_{\mbox{\tiny{min}}}^{}({\cal C}^{(N)})$ directly.
 Yet the connections between these superficially different types of questions are worth exploring further.
\smallskip

\subsubsection{Erd\H{o}s-type distance problems}

 Estimating $\nu(N)$ from below is a special case of the ``\emph{Erd\H{o}s distinct distances problem}'' \cite{Erdos} which asks for the smallest number of distinct distances that can occur in an $N$-point configuration, written as $d(N)$.
 Clearly, if ${\cal C}^{(N)}\in$ LJ$_N^{\rm gmin}$, then $\nu({\cal C}^{(N)})\geq d(N)$, and so lower bounds on $d(N)$ give lower bounds on $\nu({\cal C}^{(N)})$.
 The answer depends on the dimension of Euclidean space, but having applications to chemistry in mind, we are primarily interested in the problem for three-dimensional Euclidean space.
 However, answers obtained by considering one-, or two-dimensional Euclidean space may well yield insights into the problem in three dimensions.

 It is intuitively plausible that an $N$-point configuration exhibiting the smallest number of distinct distances among all possible $N$-point configurations is one of high symmetry, so that most if not all of the distinct distances in it occur more than once.
 This expectation is readily confirmed for small $N$.
 Recall that a regular simplex in $\Rset^D$ is a geometrical shape with $N=D+1$ vertices (locations of point particles) and $\genfrac{(}{)}{0pt}{1}{D+1}{2}$ edges that all have the same length. 
 It follows that the dimer, the equilateral trimer, and the regular tetrahedral tetramer are $D=3$ configurations with the smallest possible number of distinct pairwise distances, namely 1; i.e. $d(N)=1$ when $N\in\{2,3,4\}$ --- in $D=3$ dimensions.
 When $N>4$, then $d(N)>1$ in $D=3$ dimensions, and since the problem to determine $d(N)$ for each $N$ is NP-hard, the goal is to find good lower and upper bounds on $d(N)$.

 The trivial upper bound $d(N)\leq \genfrac{(}{)}{0pt}{1}{N}{2}$ is readily improved in one dimension by placing $N$ point particles consecutively on (a subset of) the integers $\Zset$, which yields $d(N)\leq N-1$.
 For $\Rset^2$, from a $\sqrt{N}\times\sqrt{N}$ lattice (assuming $\sqrt{N}\in\Nset$), Erd\H{o}s proved that for large $N$ one has $d(N)\leq c N/\sqrt{\log N}$ for some $c$.
 Since linear and planar configurations are embeddable in $\Rset^3$, these bounds are automatically upper bounds on $d(N)$ in $\Rset^3$.

 As for lower bounds in $\Rset^3$, it has been conjectured that when $N$ is sufficiently large, then $d(N)\geq c N^{2/3}$; the best result currently known seems to be $d(N) \geq c N^{3/5}$, proved in \cite{SolyVu}.
Hence $\nu({\cal C}^{(N)})\geq cN^{3/5}$ when $N$ is large enough, for some $c>0$.

\noindent
{\bf Remark}: {\it In 2023 Alon et al. \cite{AlonETal} proved that for generic norms any set of $N$ points in $\Rset^3$ defines at least $(1 - o(1))N$ distinct distances, thus $d(N)\geq (1 - o(1))N$.
 Since the Erd\H{o}s upper bound on planar configurations is sublinear, it follows that the Euclidean norm is not a typical norm in the sense of Alon et al. (as they themselves point out), so that their results have no bearing on the distance spectrum for LJ$_N$.}

\medskip

 Erd\H{o}s \cite{Erdos} also asked for the largest number of times that some distance between two points can occur in a cluster of $N$ points, without any constraints on whether it is the smallest one, or the largest one, or any ranking inbetween.
 The particular distance in question (i.e.~the one that occurs most often) can be taken to be the reference distance, by rescaling if necessary.
 Thus this problem is known as the ``\emph{Erd\H{o}s unit distance problem}.'' 
 We refer to this quantity as $u(N)$.
 As for the  Erd\H{o}s distinct distances problem, the answer depends on the dimension.
 It is trivial for $\Rset$ (where $u(N)=N-1$) but highly nontrivial in dimensions $D\geq 2$. 
 For the problem in the plane $\Rset^2$ the sequence is known for small $N$ and found as A186705 at \cite{OEIS}.
 For small $N$ the answer is also easy to find for $\Rset^3$; e.g.~when $N\in\{2,3,4\}$ then all possible pairs can have the same distance and $u(N) = \genfrac{(}{)}{0pt}{1}{N}{2}$, and when $N =5$ then $u(N) = \genfrac{(}{)}{0pt}{1}{N}{2}-1$.
 Yet for larger $N$ the true $N$-dependence of $u(N)$ in $\Rset^D$ will be hard to identify for any $D\geq 2$.

 Asymptotically for large $N$ some sharp estimates may be feasible.
 For the planar case Erd\H{o}s proved that $u(N) > N^{1-\epsilon}$ for any $\epsilon>0$, and he challenged his peers to find out whether there is a corresponding upper bound; as per \cite{AlonETal} the best upper bound for $N$ points in $\Rset^2$ currently known is $u(N)< C N^\frac43$, proved in the 1980s by Spencer, Szemer\'edi, and Trotter.
 For configurations in $\Rset^3$ Erd\H{o}s conjectured that $u(N) > C N^{4/3} \log \log N$ for some $C$.
 Recently Zahl \cite{Zahl} obtained the upper bound $u(N) < C N^{295/197 +\epsilon}$ in the 3D case.

\noindent
{\bf Remark}: {\it In  \cite{AlonETal} Alon et al.~also proved that for generic distances in $\Rset^3$ one has the upper bound $u(N) \leq \frac32 N\log_2 N$, and that for sufficiently large $N$ one can always find a 3D configuration $\cC(N)$ for which $u(\cC(N)) = (1- o(1))N\log_2N$.
 Again, since the Euclidean distance is not generic, the result by Alon et al.~has no bearing on the distance spectrum for equilibrium, or even minimal energy configurations in LJ$_N$.}
\smallskip

 Since for the Erd\H{o}s unit distance problem in $\Rset^3$ no demand is made on the cluster being an LJ$_N$ global energy minimum, $u(N)$ for this problem is an upper bound to any of our $C_k(N)$.
 Unfortunately, if his conjectured lower bound is true, then $u(N)$ grows superlinearly in $N$.
 To settle our conjecture about the minimal pairwise distance for the LJ$_N$ global energy minima, better estimates of the $C_k(N)$ for Lennard-Jones equilibria, or at least the global LJ$_N$ energy minima, are needed.
 
 Some estimates can be extracted from the well-known thermodynamic stability of the Lennard-Jones pair interaction, which implies that the limit, as $N\to\infty$, of the minimal Lennard-Jones energy divided by $N$ exists.
 Thus, if any particular distance, different from the one for which the Lennard-Jones pair energy yields zero, occurs superlinearly often along a sequence in the set of all LJ$_N^{\rm{min}}$, then it has to be compensated by another superlinear sequence for which the pair energy takes the opposite sign. 
 The crystal lattice results for the thermodynamic limit of the LJ$_N$ ground states reveal that this situation is impossible, because in that limit all pairwise distances that occur yield a negative LJ energy.
 This observation establishes that the minimal distance, and any other, for the configurations in LJ$_N^{\rm{min}}$, cannot occur more than $cN$ times for some constant $c$.

 To obtain an estimate for the constant $c$ we revisit systems composed of sticky hard spheres.
 We recall that $W_{\mbox{\tiny shs}}(N)$, the ground state energy achieved by optimal clusters of $N$ sticky hard spheres, by definition is the negative of the largest possible number of contacts that $N$ congruent spheres can have. 
 It is easy to see that when $N\geq 2$, then a lower bound is $-W_{\mbox{\tiny shs}}(N)\geq 2N-3$ (diagonal penny pair chain) and an upper bound is $-W_{\mbox{\tiny shs}}(N)< 12 N$ (the number of hard spheres multiplied with the maximal number of contacts a single sphere can have with congruent other spheres \cite{SchVdW}, and this overcounts the total number of contacts that can possibly exist in an $N$ body cluster of hard spheres).  
 In \cite{BezdekReid} it was proved that $-W_{\mbox{\tiny shs}}(N)\leq 6 N - 0.926 N^\frac23$ for all $N$, and the factor 6 of the linear term is sharp.
 Recalling now our discussion of the LJ gradient flows that start with such an optimal hard-sphere configuration, we suspect that $C_1(N)$ for LJ$_N^{\rm gmin}$ clusters is not larger than $-W_{\mbox{\tiny shs}}(N)$.
 Thus we expect that $C_1(N)\leq 6N$.
 \vspace{-.2truecm}

\section{The distance spectra of selected  LJ$_N^{\rm{{equ}}}$ minima }

Our approach of using the virial theorem to gain insight into the minimal pairwise distance $r_{\mbox{\tiny{min}}}(\mbox{LJ}_N^{\rm{gmin}})$ in  \LJ{N} global minima has revealed an interesting connection with the spectrum of pair distances. 
 While much is known about the spectrum of distances in regular crystal lattices, see e.g. Born's classic \cite{BornBOOKcrystals}, we are not aware of a systematic study of the spectrum of distances in LJ$_N^{\rm{gmin}}$.
 In this section we consider some numerically computed distance spectra of putative global minima for a sample of $N$ values, including  $N=923$ where the shortest distance was observed, which belongs to the centered icosahedral sequence $\{1,13,55,147,309,561,923,...\}$ (sequence A005902 at \cite{OEIS}). 
 Except for the uninteresting single atom, the other configurations with $N$ in this sequence are either Mackay icosahedra \cite{mackay62} or cuboctahedra. 
 The Mackay icosahedra yield global LJ energy minima when $N$ is not too large; with increasing $N$ the non-crystallographic packing and strain energy can make them uncompetitive with truncated crystal structures. 

\begin{figure}[hptb]
\centering
\includegraphics[width=0.615\textwidth]{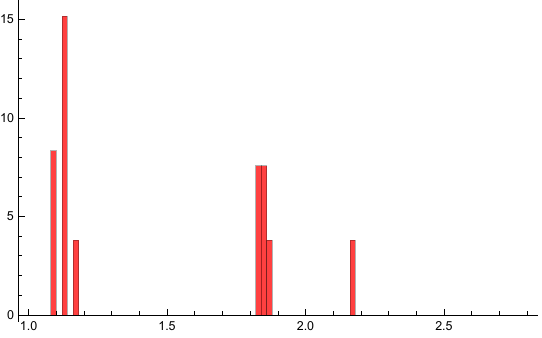}
\includegraphics[width=0.615\textwidth]{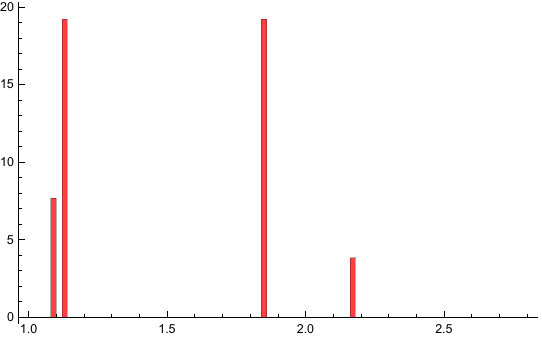}
\includegraphics[width=0.615\textwidth]{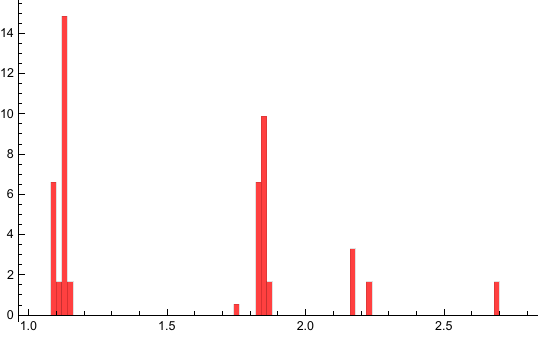}
\caption{
Distance spectra of the LJ$_N$ global minima  for $N\in\{12,13,14\}$ (top-down).
 Visibly no distance is smaller than 1. \label{fig:12}
}
\end{figure}

Figure \ref{fig:12} shows the distance spectra as  a histogram for $N\in\{12,13,14\}$, top-down in this ordering.
The bin width is 0.02 and the bin areas sum to 1.
 Clearly visible is a clustering of distances around 1.15, 
 around 1.85, and near 2.2. 
 As expected, the highly symmetrical $N=13$ Mackay icosahedron features the least number of distinct distances compared to the maximal possible number $\genfrac{(}{)}{0pt}{1}{N}{2}$, namely 4 out of 78, while the $N=12$ cluster features 7 out of 66 distinct distances, and the $N=14$ cluster 11 out of 91;
 these ratios, in decimal expansion, are: $0.1\overline{06}$,  $0.\overline{051282}$, and $0.\overline{120879}$.

 % \newpage
% \newpage

\begin{figure}[hptb]
\centering
\includegraphics[width=0.61\textwidth]{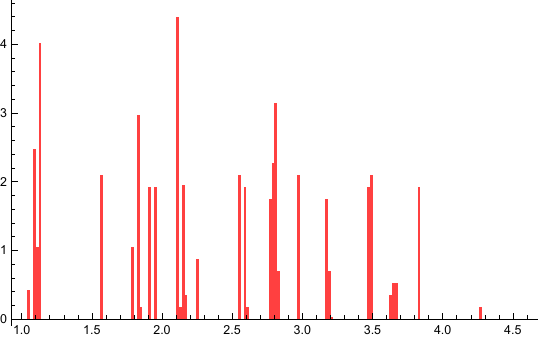}
\includegraphics[width=0.61\textwidth]{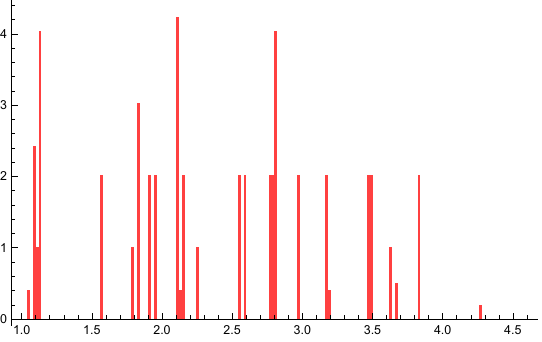}
\includegraphics[width=0.61\textwidth]{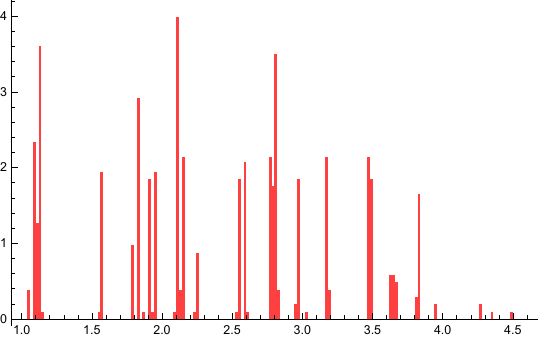}
\caption{Distance spectra of the LJ$_N$ global minima for $N\in\{54,55,56\}$ (top-down).
 Again, no distance is smaller than 1. 
 \label{fig:54}}
\end{figure}

Figure \ref{fig:54} shows the distance spectra for $N\in\{54,55,56\}$.
 The ratio of the number of distinct distances in LJ$_N$ compared to the number of distinct distances maximally possible in an $N$ body cluster is, respectively, $32/\genfrac{(}{)}{0pt}{1}{54}{2} \approx 0.022362$, $27/\genfrac{(}{)}{0pt}{1}{55}{2} = 1/55=0.0\overline{18}$, and $43/\genfrac{(}{)}{0pt}{1}{56}{2} = 0.02\overline{792207}$.
  This result is in line with the expectation that the magic number $N=55$ implies a cluster with higher symmetry than its neighbors with $N=54$ and $N=56$.

 Another feature, prominently visible for the triple $N\in\{54,55,56\}$, yet not for the triple $N\in\{12,13,14\}$, is this:
The spectrum for $N=55$ appears less ``vertically noisy'' than the one for $N=54$ or $N=56$, in the sense that the $N=55$ cluster has the largest number of distinct distances that are occupied by the same percentage number of particles.
 This will be noticeable also for the next triple near Mackay number $N=147$.
 
 The histograms for $N\in\{54,55,56\}$ are easy to compare, and so are those for $N\in\{12,13,14\}$, but the comparison of these two groups against each other is also revealing.
  Naturally, with increasing $N$ the range of distances in an LJ$_N$ cluster increases, which is  obvious by comparing the two triples of spectra. 
 Interestingly, though, in the distance spectra of the $N\in\{54,55,56\}$ clusters there also appears a notable isolated peak at $\approx 1.6$ where no such peak was visible for when $N\in\{12,13,14\}$.
 The previously visible peaks remain visible, too.

Usually, the dimension of a symmetry group is used as an absolute measure to decide which group has the higher symmetry. 
 By that measure the Mackay icosahedron with $N=55$ has the same symmetry as the one with $N=13$. 
 However, if as a \emph{relative symmetry metric} we use the reciprocal of the ratio of the actual number of distinct distances in a cluster to the maximally possible number of distances in any cluster with the same number of atoms, then the metric for  LJ$_{55}^{\rm gmin}$ is larger than for LJ$_{13}^{\rm gmin}$ by this relative metric.
 This rational symmetry measure might be useful analogously to a continuous symmetry measure \cite{zabrodskypa92,Fowler92,zabrodskypa93,zabrodskypa93a,zabrodskya95,katzenelsonza96}.

\begin{figure}[hptb]
\centering
\includegraphics[width=0.615\textwidth]{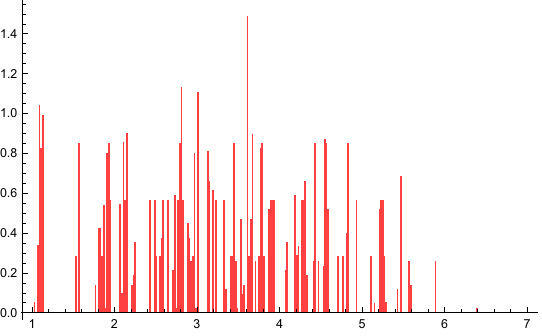}
\includegraphics[width=0.615\textwidth]{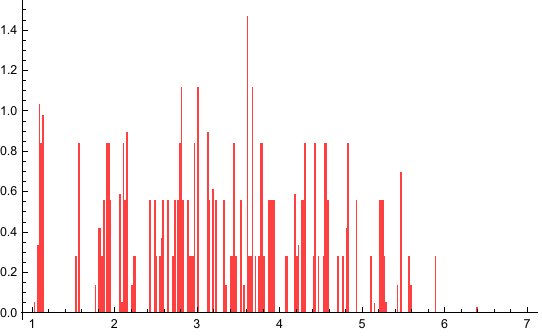}
\includegraphics[width=0.615\textwidth]{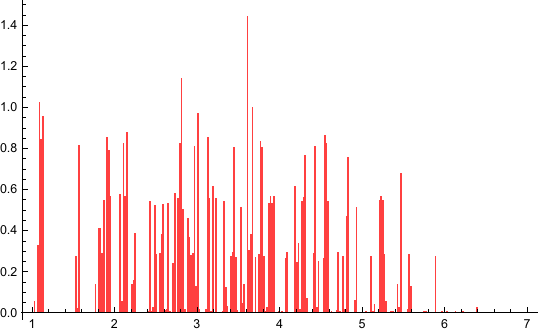}
\caption{Distance spectra of putative LJ$_{N}$ global minima for $N\in\{146,147,148\}$.
 \label{fig:146}}
\end{figure}

 Inspired by these first observations we inspect the next triple associated with the magic numbers of the Mackay icosahedra.
 The $N=147$ distance spectrum and that of its two neighbors is shown in Figure \ref{fig:146}.
  Again, not only are there fewer distinct distances in the $N=147$ spectrum than for its two adjacent neighbors, {there also are noticeably more distinct distances with equal occupation numbers than in the two spectra for the neighbours of $N=147$}.
 Also, the previously visible spectral peaks a little bit to the right of 1, near 1.6, a little bit below 2 and a little bit above 2, are all still visible. 

% \newpage

\begin{figure}[hptb]
\centering
\includegraphics[width=0.615\textwidth]{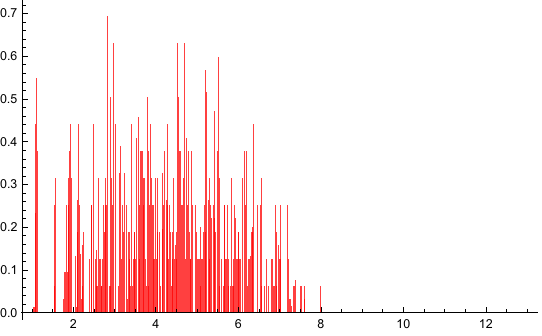}
\includegraphics[width=0.615\textwidth]{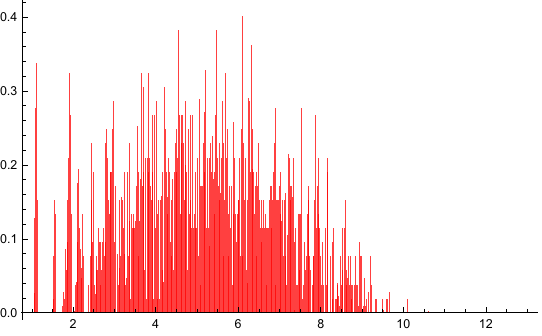}
\includegraphics[width=0.615\textwidth]{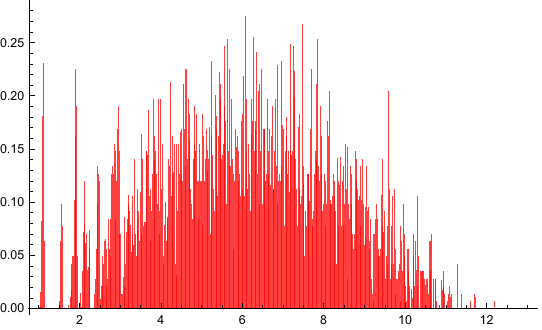}
\caption{Distance spectra of putative LJ$_{N}$ global minima for $N\in\{309,561,923\}$.
 \label{fig:309}}
\end{figure}

Going beyond $N=200$, Figure \ref{fig:309} shows the histograms for the putative LJ$_N^{\rm gmin}$ clusters with magic Mackay icosahedral numbers $N\in\{  309, 561, 923\}$:
Since the distance histogram scales with system size, and we include the full spectrum, the resolution is lower for the larger clusters.
Nevertheless the short-distance peaks a little above 1, at roughly 1.6, and a little below and above 2, are still clearly visible. 
 Also noticeable is the emergence of a broad peak feature to the right  of distance 2.2. 

\begin{figure}[hptb]
\centering
\includegraphics[width=0.61\textwidth]{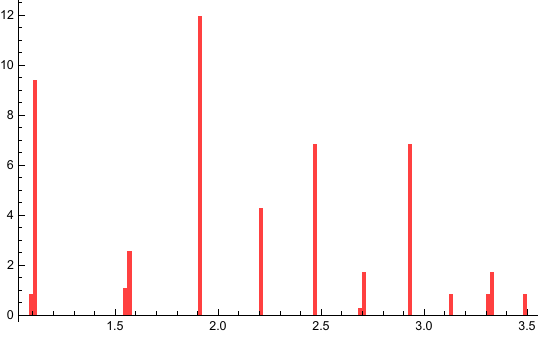}
\includegraphics[width=0.61\textwidth]{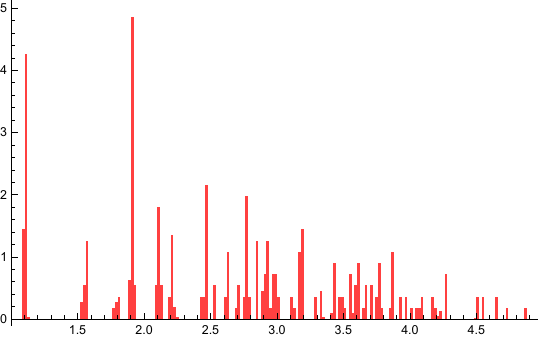}
\caption{Distance spectra for the \LJ{38} and \LJ{75} global minima (top to bottom). These normalised histograms correspond to a bin width of 0.02. 
 \label{fig:compareA}}
 \end{figure}

\newpage
{These trends are also visible in the distance spectra shown in Figure \ref{fig:compareA} for the truncated octahedral \LJ{38} and the decahedral \LJ{75} global minimum.} 

\begin{figure}[hptb]
\centering
\includegraphics[width=0.61\textwidth]{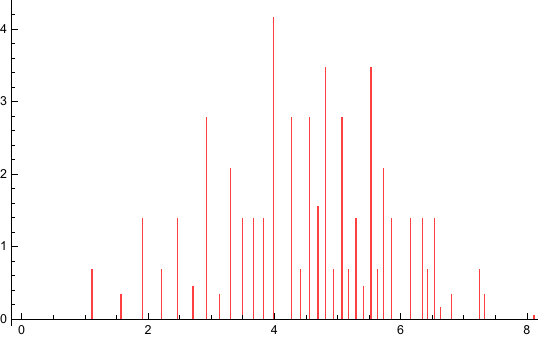}
\includegraphics[width=0.61\textwidth]{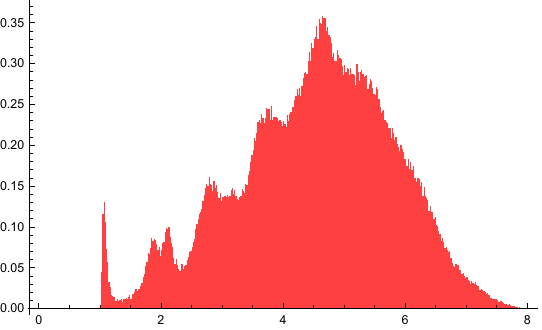}
\caption{Distance spectra for the fcc minimum and a local minimum from the liquid in a system of 864 atoms with periodic boundary conditions and number density 1.05 \cite{deSouzaW16} (top to bottom). These normalised histograms correspond to a bin width of 0.02. The horizontal scale is extended to zero for the periodic systems to highlight the first peak in the liquid configuration.
 \label{fig:compareB}}
\end{figure}

 {In Figure \ref{fig:compareB} we compare them with the corresponding results for a system with periodic boundary conditions taken from a previous study of crystallisation \cite{deSouzaW16}.}
 
The behavior of the spectrum for the global minima of finite clusters (as opposed to bulk) with increasing size is intriguing.
 As for the radial distribution function (pair correlation function) the pair distance distribution encodes a great deal of structural information, as well as the energetics, as discussed below.
 
All the histograms shown above {feature} small-distance peaks at a little above 1 and a little below and above 2, and all histograms with $N\geq 54$ also feature a peak roughly at 1.6. 
These distances can all be assigned from analysis of the underlying structures.

The pair correlation function, $g(r)$, and structure factor sampled over local minima were first used to gain new insights by Stillinger and Weber for condensed phases  \cite{stillingerw85,stillingerw85c}.
 Common neighbour analysis provides a useful way to approach the distance spectrum.
 For example, a split second peak in $g(r)$  is often associated with supercooled liquids and glasses and polytetrahedral packing \cite{Waal95,DoyeWZD03}, arising from the apex-to-apex distance in a trigonal bipyramid and approximate colinear triatomic arrangements \cite{honeycutta87,jonssona88,ClarkeJ93}. 
The smallest distance in the spectra of course corresponds to nearest-neighbour atoms, and contributions to peaks around 1.6 arise from next-nearest distances for polytetrahedral packing (1.633 for regular tetrahedra) and atoms sharing four neighbours at opposite vertices of an octahedron.
The distances a little below 2 come from the opposite vertices in triangles that share an edge, while the distance a little above 2 results from atoms in roughly collinear arrangements with a common neighbour between them. 

An interesting feature in the histograms for $N\in\{12,14\}$ is that  a unique highest peak is found very close to 1, while for all other inspected $N$ with unique highest peak it occurs at $r>2$ for the selected bin width of 0.02.
We comment further on the distance distributions and connections between structure and energetics in our Conclusions below.

%%%%%%%%%%%%%%%%%%%%%%%%%%
%%%%%%%%%%%%%%%%%%%%%%%%%%%%%%%%%%%%%%%%%%%%%%%%%%%%
\section{Conclusions}
%%%%%%%%%%%%%%%%%%%%%%%%%%%%%%%%%%%%%%%%%%%%%%%%%%%%
%%%%%%%%%%%%%%%%%%%%%%%%%%

In this article we have inquired into the minimal distance that occurs in $N$-body clusters that globally minimize the Lennard-Jones energy defined in equation (\ref{W}), or are merely stationary points.
 We have given a rigorous proof that the dimer equilibrium distance is the largest possible minimal distance, and that it occurs if and only if the $N$-body cluster is a Lennard-Jones global minimum for $N\in\{2,3,4\}$.
We have also given plausibility arguments, though no rigorous proof, for our conjecture that the minimal distance between the particles in any  Lennard-Jones $N$-body cluster global minimum is always bigger than 1 in units where the dimer equilibrium distance is $2^\frac16$.

A rigorous proof is desirable, for it would significantly reduce the search space when looking for Lennard-Jones cluster global minima. 
 The currently best known provable lower bound is about 25\% above the conjectured lower bound. 
 
  By examining the minimal pairwise distance in (mostly) numerically computed putatively energy-optimal Lennard-Jones $N$-body clusters for $2\leq N\leq 1000$ we found numerical evidence in support of our conjecture.
The plot of the minimal pairwise distance versus $N$ in addition has revealed some interesting features that suggest that the minimal pairwise distance of globally energy-minimizing LJ$_N$ clusters is a useful indicator of some structural information of the cluster. 
 We were able to map these features into centered icosahedral, icosahedral without central atom, and decahedral structures.
 It should be interesting to explore this structural information map also for larger $N$ values.
 
 Our approach to the minimial pairwise distance invoked the virial theorem for equilibrium configurations. 
 This framework provided an interesting expression for the minimal pairwise distance in an equilibrium cluster in terms of the ratio of the repulsive over the attractive part of equation (\ref{W}) for the cluster, which we then expressed in terms of features of the distance spectrum of the cluster. 
 This result in turn raised questions of the type known as Erd\H{o}s distance problems. 
 
 Inspecting the numerically computed histograms of the pair distances for some of the putative \LJ{N} global minima suggested a symmetry metric, given by the reciprocal of the number of distinct distances in the cluster relative to the maximal possible number of distinct distances, which assigns more symmetry to the $N=55$ than to the $N=13$ Mackay icosahedron, while their point groups are of course the same.

 Another feature highlighted by the distance histograms is the variation 
 in the occupation numbers of the most probable distances that feature in a distance spectrum. 
 As expected from the maximal point group symmetry, the distributions for the {Mackay} icosahedral clusters are visibly more bunched than their %immediate 
 neighbors.

The structure encodes the pair distance distribution, and hence the energy, in a unique mapping.
Connections between structure and energetics can be further analysed by defining a strain energy  in terms of the deviation of pair distances from the ideal value \cite{doyew95a,doyew96d}.
 The Morse potential \cite{morse29} has proved particularly insightful here, since it has an additional adjustable parameter compared to the Lennard-Jones representation.
 This extra degree of freedom can be interpreted in terms of the range of the pair interaction, which  can help to explain the energetic competition between different packing schemes \cite{doyew95a,doyew96d}.
 Polytetrahedral packing schemes, associated with local fivefold symmetry, can be interpreted in terms of disclinations, while quasiperiodic packings are associated with quasicrystals \cite{ShechtmanBGC84,NelsonH85,Nelson86,Steinhardt87,nelsons89}.
 The topological defects in polytetrahedral structures are associated with disclination lines \cite{nelsons83,Nelson83}, leading to Frank-Kasper phases \cite{FrankK58,FrankK59} and Kasper polyhedra.
Common neighbour \cite{honeycutta87}, Voronoi \cite{PTP.58.1079}, and bond
order \cite{steinhardtnr83} analysis have proved to be very useful in
decomposing atomic structures, while topological cluster classification
\cite{:/content/aip/journal/jcp/139/23/10.1063/1.4832897,D4SM00889H} has been
successfully applied to a variety of atomic \cite{10.1063/1.3516210,PhysRevX.7.031028,Ortlieb2023,10.1063/1.4790515} and colloidal 
\cite{TaffsMWR10,MALINS2011760} clusters.

We hope that further work along the lines we have presented here might be combined with such schemes to provide new insight into the fascinating connections between structure, energetics, and emergent physical properties.

\bigskip

\noindent
{\bf Acknowledgement}: We thank Prof.~Paddy Royall and Prof.~Jonathan Doye for helpful discussions and suggestions. 

%\newpage
%%%%%%%%%%%%%%%%%%%%%%%%%%%%%%%%%%%%%%%%%%%%%%%%%%%%%%%%%%%%%%%%%%
%%%%%%%%%%%
%%%%%%%%%%% And finally: The bibliography as per .bib file
%%%%%%%%%%%
%%%%%%%%%%%%%%%%%%%%%%%%%%%%%%%%%%%%%%%%%%%%%%%%%%%%%%%%%%%%%%%%%%
  
\bibliographystyle{thesis}
\bibliography{CLUSTERforMolPhys}

\end{document}